\documentclass{ws-ijmpa}
\usepackage[super,compress]{cite}
\usepackage{graphicx}
\usepackage{leftidx}
\begin{document}
	\markboth{AMEE KAKADIYA, CHANDNI MENAPARA and AJAY KUMAR RAI}{Mass Spectroscopy and Decay Properties of $\Xi_{cb}$, $\Xi_{bb}$ Baryons}
	
	%
	\catchline{}{}{}{}{}
	%
	
	\title{Mass Spectroscopy and Decay Properties of $\Xi_{cb}$, $\Xi_{bb}$ Baryons \footnote{Spectroscopy study of doubly heavy baryon alongwith the properties as magnetic moment, Radiative decay.}
	}
	
	\author{AMEE KAKADIYA, CHANDNI MENAPARA\footnote{chandni.menapara@gmail.com}     and AJAY KUMAR RAI}
	
	\address{Department of Physics, Sardar Valabhbhai National Institute of Technology\\
	Surat-395008, Gujarat, India
	\\
		}
	
%
	
	\maketitle
	
	\begin{history}
		\received{Day Month Year}
		\revised{Day Month Year}
	\end{history}
	
	\begin{abstract}
	Using the Hypercentral Constituent Quark Model (hCQM), the mass spectra of doubly heavy baryons $\Xi_{cb}$ and $\Xi_{bb}$ are determined. The model describes an interaction inside the baryonic system. Screened potential has been considered as confining potential with color-Coulomb potential to enumerate the masses of baryonic states.  Regge trajectories have been plotted in $(J, M^2)$ plane. The properties like magnetic moment and radiative decay width have been determined using the obtained results. 
		
	\end{abstract}
	
	\ccode{PACS numbers: 14.20.-c, 14.20.Lq, 14.20.Mr}
	
\section{Introduction}
\label{sec1}
 The study of the doubly heavy baryons is enthralling as it provides the platform to look into heavy quark symmetry and chiral dynamics\cite{QCD}. Most of the doubly heavy baryons are still unexplored experimentally \cite{review-chen}. 
Doubly heavy baryons contain two heavy quarks ($c$, $b$) and one light quark ($u$, $d$, $s$). According to strangeness, they can be divided in two families: $\Xi$ and $\Omega$. The $\Xi$ family contains only first generation light quarks ($u$, $d$) and the $\Omega$ family comprises strange quark other than two heavy quarks. One member of the doubly heavy $\Xi$ family has been discussed in our previous work\cite{Universe}, while the other two are discussed in this work.\\ 
In the doubly heavy sector, so far $\Xi_{cc}$ baryon is experimentally detected with one star status in PDG\cite{PDG}. 
In 2017, the LHCb collaboration\cite{LHCb2017} observed $\Xi_{cc}$ via $\Lambda_c^+ K^- \pi^+ \pi^-$ channel in $pp$ collision at resonance mass of 3621 MeV. A search of $\Xi_{cb}$ baryon was held by LHCb, observing the decay channel $D^0 p K^-$ at center-of-mass energy 13 TeV during $pp$ collision, but no evidence was found \cite{LHCb2020}. The mass of $\Xi_{cb}$ baryon is supposed to be in range of 6.7-7.2 $GeV/c^2$. Recently, $\Xi_{cb}$ and $\Omega_{cb}$ are detected in mass range of 6700-7300 $MeV/c^2$ via $\Lambda_c \pi$ and $\Xi_c \pi$ decay modes, during $pp$ collision with 95\% confidence level, but evidence of signal is not found \cite{LHCb2021}.  $\Xi_{bb}$ is still not experimentally detected, but there are major possibilities to be detected after experimental observation of $\Xi_{cc}$ state. The main source of doubly heavy baryons production is the direct hadronic mechanism i.e., gluon-gluon fusion mechanism and charm-bottom mechanism. Recently, the doubly heavy baryon production is proposed via Higgs decay based on non-relativistic QCD theory \cite{ma}.\\

The doubly heavy baryons investigated by many research groups with different theoretical approaches like; Hamiltonian model\cite{Yoshida}, Regge phenomenology\cite{Juhi,Juhidoubly,Wei2015,Wei2017}, Lattice QCD\cite{Rubio,Brown,Alex,Padmanath}, QCD Sum rules\cite{Aliev2012,Aliev2013,Wang2010}, the variational approach\cite{Eakins}, Relativistic approach\cite{Ebert} and many other methods\cite{Roberts,Valcarce,Giannuzzi,Gershtein,Karliner,Tang,Albertus,Martynenko,BPatel,Sun}. 
In a recent work, strong decays of low-lying doubly bottom baryons have been investigated using $\leftidx{^3}{P}{_0}$ model for the potential states to be observed by future experiments \cite{he2021}. Another article dedicated to $\Xi_{bc}$ proposed an inclusive decay channel to $\Xi_{cc}^{++}$ for LHCb Run3 \cite{qin2021}.\\
In the present work, the Hypercentral Constituent Quark Model is employed to describe the inter-quark interaction inside the doubly heavy baryons and screened potential is considered as confining potential with color-Coulomb potential which is inspired by our previous work on $\Xi_{cc}$ \cite{Universe}. Here, masses of 1S-6S, 1P-4P, 1D-4D and 1F-2F states are enumerated for $\Xi_{cb}$ and $\Xi_{bb}$ baryons. \\
The article is followed as: after the introduction, the model is described in Section \ref{sec2}. Mass spectra and Regge trajectories are presented in Section \ref{sec3}. Section \ref{sec4} deals with magnetic moment and radiative decay widths followed by conclusive remarks in Section \ref{sec5}.

\section{Theoretical Framework}
\label{sec2}
The inter-quark interaction has been described using the Hypercentral Constituent Quark Model (hCQM) wherein the internal effects are parametrized by higher constituent quark mass. The Jacobi coordinates are employed to account for the three body interactions as \cite{AKakadiya,DAE2019,ICC2019,Chandni1,Chandni2,Shahcharm,Shahtriply,ShahBBC,ShahCCB},

\begin{equation}
	\vec{\rho}=\frac{\vec{r_1}-\vec{r_2}}{\sqrt2} \hspace{0.5cm} and \hspace{0.5cm} 
	\vec{\lambda}=\frac{\vec{r_1}+\vec{r_2}-2\vec{r_3}} {\sqrt{6}}.
	\label{1}
\end{equation} 

\noindent The hypercentral coordinates (hyperradias $x$ and hyperangle $\xi$) in terms of Jacobi coordinates can be expressed as\cite{Giannini2015},

\begin{equation}
	x=\sqrt{\rho^2+\lambda^2} 
	\hspace{0.5cm} and \hspace{0.5cm}
	\xi=arctan\left(\frac{\rho}{\lambda}\right)
	\label{2}
\end{equation}

\noindent The Hamiltonian, presenting the three quark bound system is\cite{Giannini2015},

\begin{equation} 
	H=\frac{P^2}{2m} + V(x)
	\label{3}
\end{equation}

\noindent Here, $P$ is conjugate momentum and $m$ is the reduced mass of the system, which expressed as, $m=\frac{2m_\rho m_\lambda}{m_\rho + m_\lambda}$; where, $m_{\rho}=\frac{2m_1m_2}{m_1+m_2}$ and $m_{\lambda}=\frac{2m_3(m_{1}^{2}+m_{2}^{2}+m_{1}m_{2})}{(m_1+m_2)(m_1+m_2+m_3)}$ \cite{Bijkar2000}. Here,  $m_1$, $m_2$, $m_3$ are the masses of the constituent quarks: $m_u=m_d=0.344 GeV$, $m_c=1.275 GeV$, and $m_b=4.670 GeV$. As the hypercentral model itself suggests, $V(x)$ is non-relativistic interaction potential inside the baryonic system  depending only on hyperradius x (falls in two terms i.e. spin dependent ($V_{SD}$) and spin independent ($V_{SI}$) potential term \cite{Voloshin2008,Bijkar1994,AKakadiya}).

\begin{equation}
	V(x)=V_{SD}(x) + V_{SI}(x)
\end{equation} 

\noindent The expression of kinetic energy operator for the three quark system is \cite{Giannini2015},

\begin{equation} 
	\frac{P{{_x}^2}}{2m}=-\frac{\hbar^2}{2m}(\Delta_{\rho}+\Delta_{\lambda})=-\frac{\hbar^2}{2m}\left(\frac{\partial^2}{\partial x^2}+\frac{5}{x}\frac{\partial}{\partial x}+\frac{L^2(\Omega)}{x^2}\right)
	\label{4}
\end{equation}

\noindent Here, $L^2(\Omega)$ is the Grand angular operator.
\\
\noindent The screened potential is incorporated as confining potential with the color-Coulomb potential (spin independent  potential $V_{SI}(x)=V_{conf}(x) + V_{Col}(x)$) \cite{Gandhi2018,GandhiIJTP2020}. 

\begin{equation}
	V_{conf}(x)=a\left(\frac{1-e^{-{\mu} x}}{\mu}\right)  \hspace{0.5cm}  and \hspace{0.5cm} 	V_{Col}(x)= -\frac{2}{3}\frac{\alpha_s}{x}
	\label{5}
\end{equation}

\noindent where, $a$ is the string tension, the constant $\mu$(0.07) is the the screening factor, $x$ indicates the inter-quark separation and the parameter $\alpha_s$ corresponds to the strong running coupling constant \cite{ICC2019}. 

The spin dependent part of potential $V_{SD}(x)$ is \cite{Thakkar2017},
\begin{equation}
	V_{SD}(x) = V_{SS}(x)(\vec{S_{\rho}} \cdot \vec{S_{\lambda}}) + V_{\gamma S}(x)(\vec{\gamma} \cdot \vec{S}) +V_T(x) \left[ S^2 - \frac{3 (\vec{S} \cdot \vec{x}) (\vec{S} \cdot \vec{x})}{x^2} \right]
\end{equation}
which includes spin-spin, spin-orbit and tensor term respectively. The mass spectra of $\Xi_{cb}$ and $\Xi_{bb}$ baryons calculated by solving the Schr\"{o}dinger equation in Mathematica notebook\cite{Lucha1999}.

\begin{table}[tbh!]
		\tbl{Predicted masses of radial and orbital states of $\Xi_{cb}$ baryon (in GeV).}
		{\begin{tabular}{@{}ccccccccc@{}} \toprule
			 State & Present & \cite{ShahEPJC2017} & \cite{Eakins} & \cite{Roberts} & \cite{Giannuzzi} \\ \colrule
	
$1^2S_\frac{1}{2}$ & 6.915 & 6.914 & 7.014 & 7.011 & 6.904\\
$2^2S_\frac{1}{2}$ & 7.247 & 7.231 & 7.321 && 7.478\\
$3^2S_\frac{1}{2}$ & 7.481 & 7.492 &&& 7.904\\
$4^2S_\frac{1}{2}$ & 7.672 & 7.726 \\
$5^2S_\frac{1}{2}$ & 7.833 & 7.940 \\
$6^2S_\frac{1}{2}$ & 7.970\\
\hline 
$1^4S_\frac{3}{2}$ & 7.003 & 6.980 & 7.064 & 7.074 & 6.936\\
$2^4S_\frac{3}{2}$ & 7.277 & 7.256 & 7.353 & & 7.495\\
$3^4S_\frac{3}{2}$ & 7.495 & 7.505 &&& 7.917\\
$4^4S_\frac{3}{2}$ & 7.679 & 7.733 \\
$5^4S_\frac{3}{2}$ & 7.837 & 7.945 \\
$6^4S_\frac{3}{2}$ & 7.973\\
\hline 
\hline
$1^2P_\frac{1}{2}$ & 7.183 & 7.146 & 7.390\\
$1^2P_\frac{3}{2}$ & 7.179 & 7.135 & 7.394\\
$1^4P_\frac{1}{2}$ & 7.185 & 7.152 & 7.399\\
$1^4P_\frac{3}{2}$ & 7.181 & 7.141\\
$1^4P_\frac{5}{2}$ & 7.176 & 7.126\\
\hline 
$2^2P_\frac{1}{2}$ & 7.418 & 7.407 \\
$2^2P_\frac{3}{2}$ & 7.416 & 7.397 \\
$2^4P_\frac{1}{2}$ & 7.419 & 7.411 \\
$2^4P_\frac{3}{2}$ & 7.417 & 7.402 \\
$2^4P_\frac{5}{2}$ & 7.414 & 7.419 \\
\hline 
$3^2P_\frac{1}{2}$ & 7.615 & 7.642 \\
$3^2P_\frac{3}{2}$ & 7.613 & 7.634 \\
$3^4P_\frac{1}{2}$ & 7.616 & 7.646 \\
$3^4P_\frac{3}{2}$ & 7.614 & 7.638 \\
$3^4P_\frac{5}{2}$ & 7.611 & 7.653 \\
\hline
$4^2P_\frac{1}{2}$ & 7.782 & 7.859 \\
$4^2P_\frac{3}{2}$ & 7.780 & 7.852 \\
$4^4P_\frac{1}{2}$ & 7.782 & 7.863 \\
$4^4P_\frac{3}{2}$ & 7.781 & 7.856 \\
$4^4P_\frac{5}{2}$ & 7.779 & 7.869 \\
\hline
\hline
$1^2D_\frac{3}{2}$ & 7.268 & 7.303\\
$1^2D_\frac{5}{2}$ & 7.263 & 7.294\\
$1^4D_\frac{1}{2}$ & 7.274 & 7.318 \\
$1^4D_\frac{3}{2}$ & 7.270 & 7.306 & 7.324\\ 
$1^4D_\frac{5}{2}$ & 7.265 & 7.302 & 7.309\\
$1^4D_\frac{7}{2}$ & 7.259 & 7.308 & 7.292\\
\hline

\end{tabular}\label{ta1}}
{\it Table 1.}	$(${\it Continued}$)$
\end{table}

\begin{table}
\centering
{\begin{tabular}{@{}cccccccccc@{}} \toprule
 State & Present &  \cite{ShahEPJC2017} & \cite{Eakins} & \cite{Roberts} & \cite{Giannuzzi} \\ \colrule
$2^2D_\frac{3}{2}$ & 7.489 & 7.545 \\
$2^2D_\frac{5}{2}$ & 7.486 & 7.534 & 7.538\\
$2^4D_\frac{1}{2}$ & 7.493 & 7.558 & 7.579\\
$2^4D_\frac{3}{2}$ & 7.491 & 7.549 \\
$2^4D_\frac{5}{2}$ & 7.487 & 7.538 \\
$2^4D_\frac{7}{2}$ & 7.483 & 7.523 \\
\hline
$3^2D_\frac{3}{2}$ & 7.675 & 7.766 \\
$3^2D_\frac{5}{2}$ & 7.673 & 7.757 \\
$3^4D_\frac{1}{2}$ & 7.677 & 7.777 \\
$3^4D_\frac{3}{2}$ & 7.676 & 7.770 \\
$3^4D_\frac{5}{2}$ & 7.674 & 7.761 \\
$3^4D_\frac{7}{2}$ & 7.671 & 7.749 \\
\hline
$4^2D_\frac{3}{2}$ & 7.834 & 7.976 \\ 
$4^2D_\frac{5}{2}$ & 7.832 & 7.996 \\
$4^4D_\frac{1}{2}$ & 7.836 & 7.959 \\
$4^4D_\frac{3}{2}$ & 7.835 & 7.979 \\
$4^4D_\frac{5}{2}$ & 7.833 & 7.970 \\
$4^4D_\frac{7}{2}$ & 7.831 & 7.958 \\
\hline
\hline
$1^2F_\frac{5}{2}$ & 7.339 & 7.448 \\
$1^2F_\frac{7}{2}$ & 7.333 & 7.437 \\
$1^4F_\frac{3}{2}$ & 7.346 & 7.466 \\
$1^4F_\frac{5}{2}$ & 7.341 & 7.453 \\
$1^4F_\frac{7}{2}$ & 7.335 & 7.432 \\
$1^4F_\frac{9}{2}$ & 7.328 & 7.418 \\
\hline
$2^2F_\frac{5}{2}$ & 7.549 & 7.676 \\
$2^2F_\frac{7}{2}$ & 7.545 & 7.666 \\
$2^4F_\frac{3}{2}$ & 7.553 & 7.691 \\
$2^4F_\frac{5}{2}$ & 7.550 & 7.680 \\
$2^4F_\frac{7}{2}$ & 7.546 & 7.662 \\ 
$2^4F_\frac{9}{2}$ & 7.541 & 7.650 \\	
\botrule
	\end{tabular} }
\end{table}

\section{Mass Spectra and Regge Trajectories}
\label{sec3}
The mass spectra of $\Xi_{cb}$ and $\Xi_{bb}$ baryons are calculated, which include all possible hyperfine states presented in Table \ref{ta1} ranging from S to F states. As no experimental states are available for these baryons, we have compared the obtained mass with other theoretical predictions. For  $\Xi_{cb}$ baryon, our ground state is quite near to the Refs.\cite{ShahEPJC2017,Giannuzzi} but differ from Refs. \cite{Eakins,Roberts}. Further radial states are in accordance with the other predictions. $4S$ state onward, the difference of around 100 MeV can be seen which is caused by screened potential. Also in higher orbital excited states, the screening effect is visible.   \\

For $\Xi_{bb}$ baryon, our ground state prediction is matching to all other predictions with minor difference as shown in Table \ref{ta2}. And going to the higher radial states, screening effect comes in the picture with suppression of mass value. Same thing happens for the orbital states also. Mass values of 1P hyperfine states are quite matching with other comparison, but as approaching the higher orbital excited states, the effect of confining potential can be seen.  \\

\begin{table}
	\tbl{Predicted masses of radial and orbital states of $\Xi_{bb}$ baryon (in GeV).}
	{\begin{tabular}{@{}cccccccccc@{}} \toprule
		State & Present  & \cite{ShahEPJC2017} & \cite{Yoshida} & \cite{Eakins} & \cite{Roberts} & \cite{Valcarce} & \cite{Ebert} & \cite{Juhi} \\ \colrule
		$1^2S_\frac{1}{2}$ & 10.221 & 10.317 & 10.314 & 10.322 & 10.340 & 10.189 & 10.202 & 10.230 \\
		$2^2S_\frac{1}{2}$ & 10.525 & 10.605 & 10.571 & 10.551 & 10.576 & 10.482 & 10.441\\
		$3^2S_\frac{1}{2}$ & 10.749 & 10.851 & 10.612 &&&& 10.630\\
		$4^2S_\frac{1}{2}$ & 10.940 & 11.073 &&&&& 10.812\\
		$5^2S_\frac{1}{2}$ & 11.107 & 11.278\\
		$6^2S_\frac{1}{2}$ & 11.255 & \\
		\hline 
		$1^4S_\frac{3}{2}$ & 10.261 & 10.340 & 10.339 & 10.352 & 10.367 & 10.218 & 10.237 & 10.333\\
		$2^4S_\frac{3}{2}$ & 10.540 & 10.613 & 10.592 & 10.574 & 10.578 & 10.501 & 10.482\\
		$3^4S_\frac{3}{2}$ & 10.756 & 10.855 & 10.593 &&&& 10.673\\
		$4^4S_\frac{3}{2}$ & 10.943 & 11.075 &&&&& 10.856\\
		$5^4S_\frac{3}{2}$ & 11.109 & 11.280 \\
		$6^4S_\frac{3}{2}$ & 11.256 \\
		\hline 
		\hline
		$1^2P_\frac{1}{2}$ & 10.458 & 10.507 & 10.476 & 10.691 & 10.493 & 10.406 & 10.368 & 10.499\\
		$1^2P_\frac{3}{2}$ & 10.456 & 10.502 & 10.476 & 10.692 & 10.495 & & 10.408 & 10.615\\
		$1^4P_\frac{1}{2}$ & 10.459 & 10.510\\
		$1^4P_\frac{3}{2}$ & 10.457 & 10.505\\
		$1^4P_\frac{5}{2}$ & 10.455 & 10.514 & 10.759 & 10.695\\
		\hline 
		$2^2P_\frac{1}{2}$ & 10.686 & 10.758 & 10.703 && 10.710 & 10.612 & 10.563\\
		$2^2P_\frac{3}{2}$ & 10.685 & 10.754 & 10.704 && 10.713 && 10.607\\
		$2^4P_\frac{1}{2}$ & 10.687 & 10.760\\
		$2^4P_\frac{3}{2}$ & 10.685 & 10.756\\
		$2^4P_\frac{5}{2}$ & 10.684 & 10.751 & 10.773 && 10.713\\
		\hline 
		$3^2P_\frac{1}{2}$ & 10.883 & 10.985\\
		$3^2P_\frac{3}{2}$ & 10.882 & 10.981\\
		$3^4P_\frac{1}{2}$ & 10.883 & 10.987\\
		$3^4P_\frac{3}{2}$ & 10.882 & 10.983\\
		$3^4P_\frac{5}{2}$ & 10.881 & 10.978\\
		\hline
		$4^2P_\frac{1}{2}$ & 11.055 & 11.194\\
		$4^2P_\frac{3}{2}$ & 11.055 & 11.191\\
		$4^4P_\frac{1}{2}$ & 11.056 & 11.196\\
		$4^4P_\frac{3}{2}$ & 11.055 & 11.193\\
		$4^4P_\frac{5}{2}$ & 11.054 & 11.188\\
		\hline
		\hline
		$1^2D_\frac{3}{2}$ & 10.549 & 10.658 &&&&&&10.761\\
		$1^2D_\frac{5}{2}$ & 10.547 & 10.652 &&&&&&10.889\\
		$1^4D_\frac{1}{2}$ & 10.551 & 10.665\\ 
		$1^4D_\frac{3}{2}$ & 10.550 & 10.660\\
		$1^4D_\frac{5}{2}$ & 10.547 & 10.654\\
		$1^4D_\frac{7}{2}$ & 10.545 & 10.647\\
		\hline
		
	\end{tabular}\label{ta2}}
	{\it Table 2.}	$(${\it Continued}$)$
\end{table}

\begin{table}
\centering
{\begin{tabular}{@{}cccccccccc@{}} \toprule
State & Present  & \cite{ShahEPJC2017} & \cite{Yoshida} & \cite{Eakins} & \cite{Roberts} & \cite{Valcarce} & \cite{Ebert} & \cite{Juhi}  \\ \colrule		
		$2^2D_\frac{3}{2}$ & 10.765 & 10.891\\
		$2^2D_\frac{5}{2}$ & 10.763 & 10.888\\
		$2^4D_\frac{1}{2}$ & 10.767 & 10.897\\
		$2^4D_\frac{3}{2}$ & 10.766 & 10.893\\
		$2^4D_\frac{5}{2}$ & 10.764 & 10.886\\
		$2^4D_\frac{7}{2}$ & 10.762 & 10.881\\
		\hline
		$3^2D_\frac{3}{2}$ & 10.953 & 11.105\\
		$3^2D_\frac{5}{2}$ & 10.952 & 11.101\\
		$3^4D_\frac{1}{2}$ & 10.954 & 11.109\\
		$3^4D_\frac{3}{2}$ & 10.953 & 11.107\\
		$3^4D_\frac{5}{2}$ & 10.952 & 11.103\\
		$3^4D_\frac{7}{2}$ & 10.951 & 11.097\\
		\hline
		$4^2D_\frac{3}{2}$ & 11.119 & 11.306\\
		$4^2D_\frac{5}{2}$ & 11.118 & 11.303\\
		$4^4D_\frac{1}{2}$ & 11.120 & 11.310\\
		$4^4D_\frac{3}{2}$ & 11.119 & 11.307\\
		$4^4D_\frac{5}{2}$ & 11.118 & 11.302\\
		$4^4D_\frac{7}{2}$ & 11.117 & 11.298\\
		\hline
		\hline
		$1^2F_\frac{5}{2}$ & 10.626 & 10.797 &&&&&& 11.017\\
		$1^2F_\frac{7}{2}$ & 10.623 & 10.799 &&&&&& 11.157\\
		$1^4F_\frac{3}{2}$ & 10.629 & 10.805\\
		$1^4F_\frac{5}{2}$ & 10.627 & 10.790\\
		$1^4F_\frac{7}{2}$ & 10.624 & 10.792\\
		$1^4F_\frac{9}{2}$ & 10.620 & 10.784\\
		\hline
		$2^2F_\frac{5}{2}$ & 10.833 & 11.016\\
		$2^2F_\frac{7}{2}$ & 10.831 & 11.010\\
		$2^4F_\frac{3}{2}$ & 10.835 & 11.023\\
		$2^4F_\frac{5}{2}$ & 10.834 & 11.018\\
		$2^4F_\frac{7}{2}$ & 10.832 & 11.012\\
		$2^4F_\frac{9}{2}$ & 10.829 & 11.005\\
 \botrule	
	\end{tabular} }
\end{table}

	

	\begin{figure}
		\begin{minipage}{.45\textwidth}
			\includegraphics[scale=0.3]{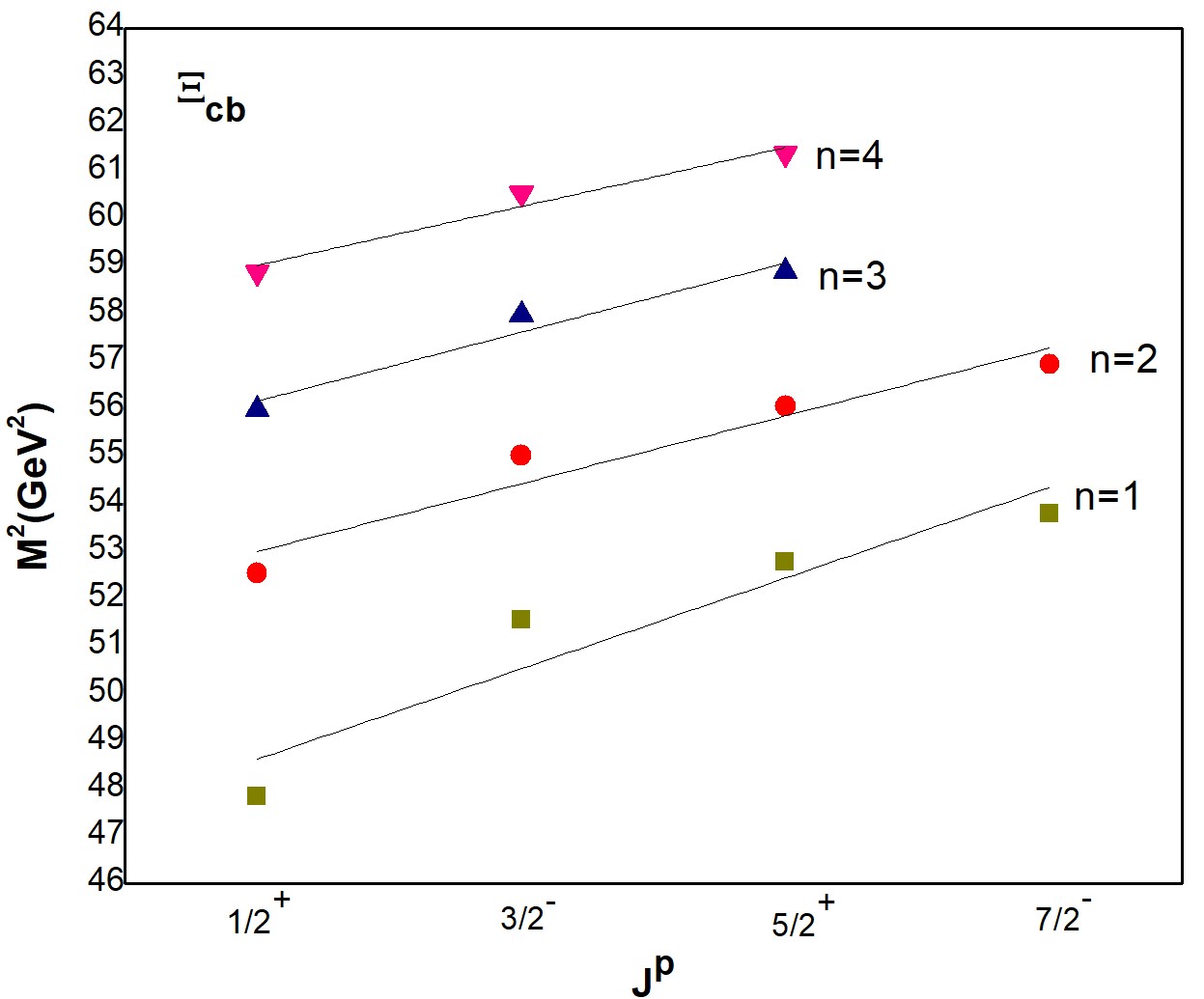}
		\caption{The $M^2 \rightarrow J$ Regge trajectory of $\Xi_{cb}$ baryon with natural parity.}
		\label{fig 1}
		\end{minipage}
		\hfill
		\begin{minipage}{.5\textwidth}
			
			\includegraphics[scale=0.3]{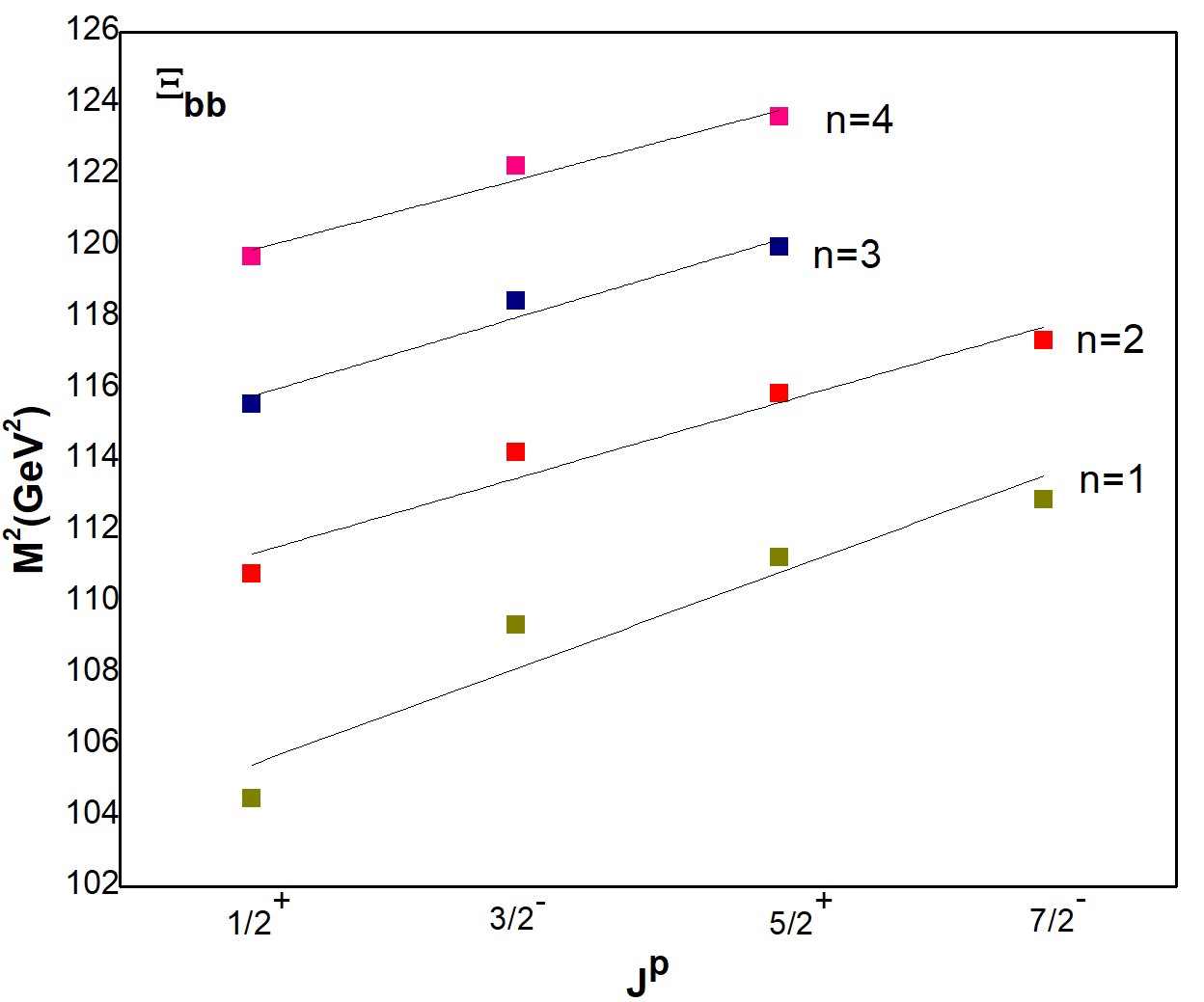}
			\caption{The $M^2 \rightarrow J$ Regge trajectory of $\Xi_{bb}$ baryon with natural parity.}
			\label{fig 2}
		\end{minipage}
	\end{figure}
 Regge trajectories are plotted in ($J, M^2$) plane for $\Xi_{cb}$ and $\Xi_{bb}$ baryons using calculated spectroscopic data. As the Regge trajectories known to be linear for baryons, it helps to justify the calculated mass spectra and also helps to define quantum number to the particular resonance mass. The Regge trajectories have been plotted for all natural parity states ($J^p = \frac{1}{2}^+, \frac{3}{2}^-, \frac{5}{2}^+ $ and $ \frac{7}{2}^-$) using the equation,
 \begin{equation}
 	J=\alpha M^2 + \alpha_0
 \end{equation}

Here, $J$ is total angular quantum number for a particular baryonic state, $M^2$ is the square of mass of the baryonic state, $\alpha$ is the slope of Regge line and $\alpha_0$ is an intercept of Regge line on $y$-axis.

The Regge plots for $\Xi_{cb}$ and $\Xi_{bb}$ baryons are shown in figure \ref{fig 1} and \ref{fig 2}. Here, our Regge trajectories seem linear but also look slightly contracting for higher excited states, which shows the screening effect caused by the confining potential used in theoretical framework.  	
	

\section{Magnetic moment}
\label{sec4}
\begin{table}[h]
	\tbl{Magnetic moment of spin 1/2 and 3/2 of $\Xi_{cb}$ and $\Xi_{bb}$ baryons in $\mu_N$.}
	{\begin{tabular}{@{}ccccccccc@{}} \toprule
			Baryon& $J^p$ & Expression & (Present) & \cite{ShahEPJC2017} & \cite{BPatel} & \cite{Bernotas} & \cite{Dhir2009} & \cite{Albertus} \\	\colrule
			$\Xi_{cb}^+$ & $\frac{1}{2}^+$ &  $\frac{2}{3}\mu_c+\frac{2}{3}\mu_b-\frac{1}{3}\mu_u$  & -0.304 & -0.204 & -0.400 & -0.236 & -0.369 & -0.475 \\
			
			$\Xi_{cb}^0$ & $\frac{1}{2}^+$ & $\frac{2}{3}\mu_c+\frac{2}{3}\mu_b-\frac{1}{3}\mu_d$  & 0.528 & 0.354 & 0.476 & 0.068 & 0.480 & 0.518\\
			
			$\Xi_{bb}^0$ & $\frac{1}{2}^+$ & $\frac{4}{3}\mu_b-\frac{1}{3}\mu_u$ & -0.669 & -0.663 & -0.656 & -0.432 & -0.630 & -0.742\\
			
			$\Xi_{bb}^-$ & $\frac{1}{2}^+$ & $\frac{4}{3}\mu_b-\frac{1}{3}\mu_d$ & 0.198 & 0.196 & 0.190 & 0.086 & 0.215 & 0.251\\
			
			$\Xi_{cb}^{*+}$ & $\frac{3}{2}^+$ & $\mu_c+ \mu_b+ \mu_u$ & 2.040 & 1.562 & 2.052 & 1.414 & 2.022 & 2.270 \\
			
			$\Xi_{cb}^{*0}$ & $\frac{3}{2}^+$ & $\mu_c+ \mu_b+ \mu_d$ & -0.422 & -0.372 & -0.567 & -0.257 & -0.508 & -0.712\\
			
			$\Xi_{bb}^{*0}$ & $\frac{3}{2}^+$ & $2\mu_b+ \mu_u$ & 1.619 & -1.607 & 1.576 & 0.916 & 1.507 & 1.870 \\
			
			$\Xi_{bb}^{*-}$ & $\frac{3}{2}^+$ & $2\mu_b+ \mu_d$ & -0.970 & -1.737 & -0.951 & -0.652 & -1.029 & -0.522\\

			\botrule	
		\end{tabular} \label{ta3}}
\end{table}

The magnetic moment of the baryon is an important intrinsic property. The magnetic moment expression of baryon can be derive by operating the expectation value equation given below \cite{GandhiDecay},
\begin{equation}
	\mu_B=\sum_q \langle \Phi_{sf}|\hat{\mu}_{qz}|\Phi_{sf} \rangle;  \quad \quad q=u, d, b
\end{equation}
Here, $\Phi_{sf}$ represents the spin-flavour wave-function of the baryon and $\hat{\mu}_{qz}$ is the magnetic moment operator. 
\noindent The magnetic moment of individual quark is given by \cite{GandhiDecay},
\begin{equation}
	\mu_q=\frac{e_q}{2m_q^{eff}}\cdot \sigma_q
\end{equation}
where, $e_q$ and $\sigma_q$ are charge and spin of the individual constituent quark of the baryonic system respectively and $m_q^{eff}$ is the effective mass of constituent quark which is expressed as given below \cite{GandhiDecay}:
\begin{equation}
	m_q^{eff}=m_q \left(1+\frac{\langle H \rangle}{\sum_q m_q}\right)
\end{equation}
Here, the Hamiltonian $\langle H \rangle$ is expressed as the difference of predicted mass(in experiment, measured mass) and total of the individual constituent masses of the baryon ($\langle H \rangle=M-\sum_q m_q$; where, $M$ is predicted mass of the particular state). And $m_q^{eff}$ is the mass of bounded quark inside the baryon with consideration of the interaction with other two quarks. 
The magnetic moment of the baryonic systems along with their respective spin-flavor wave-function are shown in the Table \ref{ta3} as well as the results are compared with other approaches.

\subsection{Radiative decay}
Transition magnetic moment and thus radiative decay width play an important role in the understanding of electromagnetic interactions and thus internal properties of a baryon.
The radiative decay width is expressed as \cite{GandhiDecay},
\begin{equation}
	\Gamma=\frac{k^3}{4\pi}\frac{2}{2J+1}\frac{e^2}{2m_p^2}\mu_{B\rightarrow B'}^2
\end{equation}
where, $k$ is photon energy, $J$ is total angular momentum of the initial baryonic state, $m_p$ is the mass of proton (in MeV) and $\mu_b$ transition magnetic moment for the particular radiative decay.

The radiative transition magnetic moment can be calculated by the sandwiching spin-flavour wave functions of initial baryon state($B$) and final baryon state($B'$) with $z$ component of magnetic moment operator, which is expressed as below:
\begin{equation}
	\mu_{B\rightarrow B'}=\langle \Phi_B|\mu_{B\rightarrow B'}|\Phi_{B'}\rangle    
\end{equation}
The spin-flavour wave function of initial baryon($\Phi_B$) state and final baryon state($\Phi_B'$) can be determined as described in  \cite{Majethiya2009}.
\begin{table}[h]
	\tbl{Radiavtive Transition magnetic moment and radiative decay width.}
	{\begin{tabular}{@{}ccccccccc@{}} \toprule	
			
Transition & Expression & (Present) &  Transition magnetic &Radiative decay width&  \\
&  &(in $\mu_N$) & moment \cite{Li} & (Present) (in keV) \\	  \colrule
$\Xi_{cb}^{*+}\rightarrow \Xi_{cb}^+ \gamma$ & $\frac{\sqrt{2}}{3}(\mu_c+ \mu_b-2\mu_u)$ & -1.340 &-1.61& 12.078\\
$\Xi_{cb}^{*0}\rightarrow \Xi_{cb}^0 \gamma$ & $\frac{\sqrt{2}}{3}(\mu_c+ \mu_b-2\mu_d)$ & 0.946 &1.02& 12.079\\
$\Xi_{bb}^{*0}\rightarrow \Xi_{bb}^0 \gamma$ & $\frac{2\sqrt{2}}{3}(\mu_b-\mu_u)$ & -1.709 &-1.81& 1.148 \\
$\Xi_{bb}^{*-}\rightarrow \Xi_{bb}^- \gamma$ & $\frac{2\sqrt{2}}{3}(\mu_b-\mu_d)$ & 0.737 &0.81& 1.149\\	
 \botrule	
\end{tabular} \label{ta4}}
\end{table}

The radiative transition decay widths are shown in Table \ref{ta4}, which is calculated by using transition magnetic moment for the particular transition.

\section{Conclusion}
\label{sec5}
The mass spectra of radial and orbital states of doubly heavy baryons are predicted using Hypercentral Constituent Quark Model (hCQM),by employing screened potential as a confining potential with color-Coulomb potential and compared with references with different theoretical approaches \cite{ShahEPJC2017,Yoshida,Eakins,Roberts,Valcarce,Ebert,Giannuzzi}. As depicted from the tables with comparison, the present results are suppressed especially at higher radial and orbital when compared to linear confining potential appearing in previous work \cite{ShahEPJC2017}. In case of \cite{Roberts} studying using spin-multiplet in HQET, the radial state for spin $\frac{1}{2}$ and $\frac{3}{2}$ is vary within 100 MeV. For $\Xi_{bc}$, only Eakins et al. \cite{Eakins}, provided few orbital state including $\frac{5}{2}$, $\frac{7}{2}$ through systematics and symmetries approach which are higher than our results. Our results are underpredicted compared to another non-relativistic approach discussed in ref \cite{Yoshida} as screening effect appears for higher spin configuration. It is noteworthy that our results are on higher side when compared with relativistic approach \cite{Ebert} and Faddevv approach \cite{Valcarce} available for S and P spin states. For F-state, the mass difference is quite large when compared with Regge phenomenology approach \cite{Juhi}.

From Regge trajectories, further excited states masses and quantum number for particular resonance masses can be predicted. Here, the Regge lines seems slightly contracting for higher excited states caused by the screening effect at higher excited states.

Radiative decay of doubly heavy baryons are studied and radiative decay widths are calculated for spin state transition of $\Xi_{cb}$ and $\Xi_{bb}$ baryons. The ground state magnetic moment and transition magnetic moment are also enumerated by using the calculated spectroscopic data. This study is expected to be useful for future experimental searches of doubly heavy baryons.

\section*{Acknowledgement}
The authors are thankful to the organizers of 10th International Conference on New Frontiers in Physics (ICNFP 2021) for giving the opportunity to present our work. Also, Ms. Chandni Menapara is thankful for support under DST-INSPIRE Fellowship.
	

\end{document}